\documentclass{article}
\usepackage{spconf,amsmath,graphicx}
\usepackage{amsfonts}
\usepackage{booktabs}
\usepackage[T1]{fontenc}

\title{Source separation with weakly labelled data: An approach to computational auditory scene analysis}
\name{Qiuqiang Kong$^{1*}$\thanks{* Part of this work was undertaken while Qiuqiang Kong was at University of Surrey.}, Yuxuan Wang$^{1}$, Xuchen Song$^{1}$, Yin Cao$^{2}$, Wenwu Wang$^{2}$, Mark D. Plumbley$^{2}$}
\address{$^1$ ByteDance AI Lab \\ $^{1}$\{kongqiuqiang, wangyuxuan.11, xuchen.song\}@bytedance.com \\ $^2$ Centre for Vision, Speech and Signal Processing (CVSSP), University of Surrey, UK \\ $^{2}$ \{yin.cao, w.wang, m.plumbley\}@surrey.ac.uk}
\begin{document}
%
\maketitle
\begin{abstract}
Source separation is the task to separate an audio recording into individual sound sources. Source separation is fundamental for computational auditory scene analysis. Previous work on source separation has focused on separating particular sound classes such as speech and music. Many of previous work require mixture and clean source pairs for training. In this work, we propose a source separation framework trained with weakly labelled data. Weakly labelled data only contains the tags of an audio clip, without the occurrence time of sound events. We first train a sound event detection system with AudioSet. The trained sound event detection system is used to detect segments that are mostly like to contain a target sound event. Then a regression is learnt from a mixture of two randomly selected segments to a target segment conditioned on the audio tagging prediction of the target segment. Our proposed system can separate 527 kinds of sound classes from AudioSet within a single system. A U-Net is adopted for the separation system and achieves an average SDR of 5.67 dB over 527 sound classes in AudioSet. 
 
\end{abstract}
\section{Introduction}
\label{sec:introduction}
Source separation is the task to separate sound sources in an audio recording. Source separation is fundamental for computational auditory scene analysis (CASA) \cite{wang2008computational}. In essence, CASA systems are machine listening systems that aim to separate mixtures of sound sources in the same way that human listeners do. The goal of a CASA system is to detect and separate sound sources from an audio recording. CASA is a challenging problem because there are large number of different types of sound events in the world. Sound events can occur simultaneously which leads to the well known cocktail party problem. 

Source separation has been researched for several years. Early work on source separation includes non-negative matrix factorization \cite{virtanen2007monaural, bruna2015source} by learning dictionaries for different sound sources. Sparse representations have been proposed for music separation in \cite{plumbley2009sparse}. An unsupervised method using average harmonic structure modeling was proposed for music source separation in \cite{duan2008unsupervised}. Recently neural network based methods have been used for source separation which learn a regression from the mixture of sources to an individual source. Neural network based methods include fully connected neural networks \cite{xu2015regression}, convolutional neural networks (CNNs) \cite{grais2017single, takahashi2017multi, jansson2017singing} and recurrent neural networks (RNNs) \cite{huang2015joint, uhlich2017improving}. Weakly supervised source separation methods include \cite{zhang2017weakly, stoller2018adversarial, stowell2015denoising, karamatli2019audio, kong2019sound, kavalerov2019universal}

However, most of the source separation systems are designed to separate specific sound classes such as speech or music. In contrast, CASA is a more challenging task that requires to separate all sound sources in the world. First, there can be hundreds of sound classes in the real world that increases the difficulty to separate all of these sound sources. Second, previous source separation systems require mixture and clean source pairs for training. For example, to separate vocals and accompaniment researchers need to collect clean vocals and accompaniment sources. However, collecting the clean sources is time consuming so the size of dataset is often limited. In addition, it is impractical to collect clean sources for a large amount of sound classes in the real world for CASA such as sounds of nature and sounds of animals. On the other hand, AudioSet is a large-scale dataset contains 5000 hours of data with 527 sound classes. AudioSet is a weakly labelled dataset, that is, only the tags of an audio clip are provided, without the occurrence time of the sound classes, neither the clean sources.

Previous work has not investigated training a source separation system without clean sources. Human beings can learn to identify and separate a sound source even if they have not heard the clean source. Humans can detect sound events around them and learn what the sounds are even if the sound events are mixed with other sounds, which is a usual case in the real world. 

In this work, we propose a source separation framework to separate a large amount of sound classes trained with weakly labelled data only. As far as we know, this is the first attempt to solve the general source separation problem for a large amount of sound classes trained with weakly labelled data. First, we train a sound event detection (SED) system on AudioSet. Then the SED system is used to detect anchor segments in an audio clip that is most likely contain a sound event. Then a regression is learned from the mixture of two randomly selected anchor segments to the target anchor segment using the audio tagging prediction of the target anchor segment as condition.

This paper is organized as follows. Section \ref{section:regression_based_ss} introduces the regression based source separation method. Section \ref{section:ss_audioset} proposes the source separation framework of a large amount of sound classes trained with weakly labelled data. Section \ref{section:experiment} shows the experiments of source separation of AudioSet. Section \ref{section:conclusion} concludes this work. 

\section{Regression based source separation}
\label{section:regression_based_ss}
Neural network based regression methods have been used to solve music separation and speech separation in \cite{xu2015regression, grais2017single, takahashi2017multi, huang2015joint, uhlich2017improving}. Regression based source separation methods learn a mapping from a mixture of sources to a target source to be separated. We denote the individual sources as $ s_{1}, .., s_{K} $, where $ K $ is the number of sound sources and each source $ s $ is a time domain signal. The mixture is denoted as $ x $. Previous source separation systems build regressions for each individual source $ s_{k} $ : 

\begin{equation} \label{eq1}
f(x) \mapsto s_{k}. 
\end{equation}

\noindent For speech enhancement, $ s_{k} $ can be clean speech. For music source separation, $ s_{k} $ can be vocal or accompanies. In this work, we build $ f $ in the time-frequency (T-F) domain \cite{xu2015regression, uhlich2017improving}. The mixture $ x $ and source $ s_{k} $ are transformed to spectrum denoted as $ X $ and $ S_{k} $ using short time Fourier transform (STFT). The magnitude and phase of $ X $ are denoted as $ |X| $ and $ e^{j \angle{X}} $ where $ X = |X| e^{j \angle{X}} $. The magnitude $ |X| $ is called spectrogram. A neural network $ g $ is built to regress from a mixture spectrogram $ |X| $ to an estimated source spectrogram $ |\hat{S}| $. Then, the phase from the mixture is used to recover the STFT of the estimated source $ \hat{S} = |\hat{S}| e^{j \angle{X}} $. Finally an inverse STFT is applied on $ \hat{S} $ to obtain the separated source $ \hat{s} $. 

In this work we model the neural network $ g $ with a U-Net \cite{jansson2017singing}. U-Net is a variation of CNN that consists of several encoding and decoding blocks modeled by convolutional layers. Each encoding block consists of two convolutional layers and a downsampling layer that halves the size of the feature maps which encode a spectrogram to a small and deep representation. Each decoding block consists of two convolutional layers and a transposed convolutional layer. The U-Net adds additional skip connections between blocks at the same hierarchical level in the encoder and decoder. This allows low-level information to flow directly from the high resolution input to the high-resolution output.

\section{Source separation with weakly labelled data}\label{section:ss_audioset}
In this section we propose a source separation framework trained with weakly labelled AudioSet. 
\subsection{Sound event detection}\label{section:sed}
Each audio clip in AudioSet is only labelled with the tags of the sound classes without their occurrence time. To solve this problem, we train a SED system \cite{xu2018large, kong2019panns} using weakly labelled AudioSet. The trained SED system is used to detect when a sound event happens in an audio clip. We choose the segment that most likely contains a sound event according to the SED prediction and refer to the selected segment as an \textit{anchor segment} for a sound class in the audio clip.

To train the SED system with weakly labelled data, a log mel spectrogram is used as feature for an audio clip. Then a CNN is applied on the log mel spectrogram \cite{xu2018large}. To predict the presence probability of sound events over time, we apply time distributed fully connected layers followed by sigmoid non-linearity to predict the presence probability of sound events over time. We denote the probability over time as $ O(t) \in [0, 1]^{K}, t=1, ..., T $ where $ T $ is the number of time steps in the distributed fully connected layer. In training, the time distributed probability $ O(t) $ is pooled to a clipwise prediction $ \hat{y} \in [0, 1]^{K} $ by a maximum aggregation function. The segment selected by the maximum aggregation has high confidence of containing the sound event to be detected \cite{xu2018large}. Binary crossentropy $ - \sum_{k=1}^{K} [ y_{k} \text{ln} \hat{y}_{k} + (1 - y_{k}) \text{ln}(1 - \hat{y}_{k}) ]$ is used for training the SED system. 

In inference, an audio clip from AudioSet is used as input to the SED system. The time distributed prediction of sound events is obtained from $ O(t) $. For a sound class $ k $, we select the time step with the highest presence probability $ \tau = \underset{t}{\text{argmax}} \ O(t)_{k} $. The anchor segment is obtained by selecting a waveform segment with $ \tau $ as centre. The trained SED system is from \cite{kong2019panns}\footnote{https://github.com/qiuqiangkong/audioset\textunderscore tagging\textunderscore cnn}. 

\subsection{Source separation}
One way to build the source separation system is to build one system for each individual sound event as mentioned in Section \ref{section:regression_based_ss}. However this is impractical because the number of source separation systems will increase linearly with the number of sources to be separated. In this section we propose a unified framework that can separate all sources in one system. 

To begin with, we randomly select two sound classes from AudioSet. For each sound class, an audio clip containing the sound class is selected. The SED system is applied to an audio clip to select the anchor segment that contains the sound event. The selected anchor segments for the two sound classes are denoted as $ s_{1} $ and $ s_{2} $, respectively. In addition to the anchor segment, a condition vector is used as an extra input to control what source to separate. For example, we denote the condition vector as $ c_{j} $ for source $ s_{j} $. Then the proposed source separation system can be described as:
\begin{equation} \label{eq:conditional_ss}
f(s_{1} + s_{2}, c_{j}) \mapsto s_{j}
\end{equation}
\noindent where $ j \in \{1, 2\} $. Equation (\ref{eq:conditional_ss}) shows that the separated audio depends on both the input mixture and the condition vector. The condition vector $ c_{j} $ controls what source to be separated.

One challenge of source separation is the aforementioned separation systems in Section \ref{section:regression_based_ss} require both mixture and clean source for training. A clean source $ s_{j} $ only contain sound of one class. However, AudioSet only provides weakly labelled audio clips and there is no information when a sound event occurs. In addition, AudioSet usually contain multiple sound classes in one audio clip so is not clean. The anchor segments selected by the SED system may contain both the selected sound class and other polyphonic sound classes. 

\begin{figure}[t]
  \centering
  \centerline{\includegraphics[width=\columnwidth]{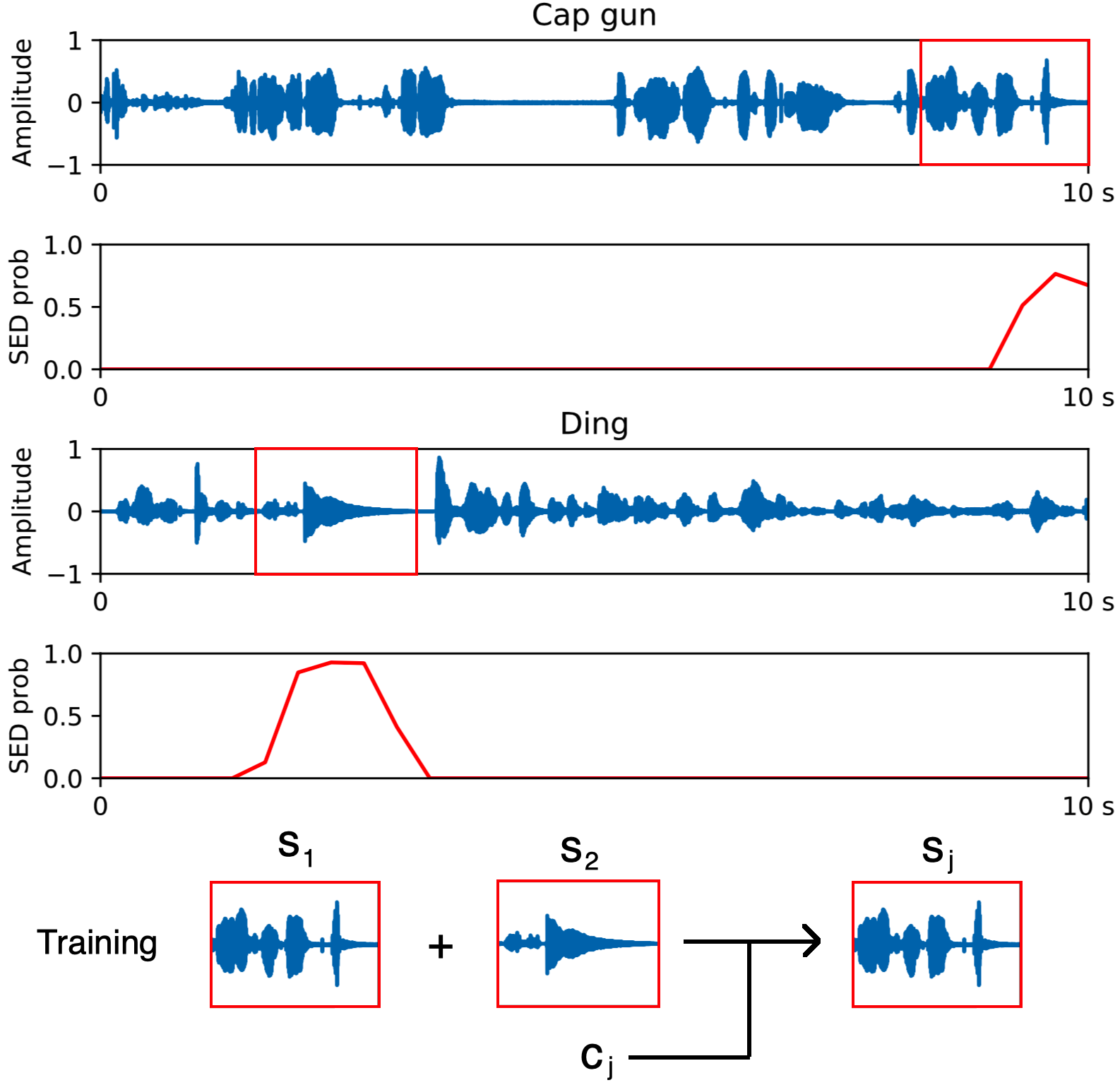}}
  \caption{The first and third rows are waveforms of two audio clips. The red rectangles are selected anchor segments described in Section \ref{section:sed}. The second and fourth rows are the SED predictions of two sound classes. }
  \label{fig:framework}
\end{figure}

We propose that training a source separation system does not necessarily require clean sources. The solution is by setting the conditional vector $ c_{j} $ to reflect the target source $ s_{j} $ to be separated. We set $ c_{j} \in [0, 1]^{K} $ that reflects the presence probability of all $ K $ sound classes in $ s_{j} $. If an anchor segment $ s_{j} $ only contains clean source for the $k$-th sound class, then $ c_{j} $ should be one hot encoding of the $k$-th sound class. On the other hand, if an anchor segment $ s_{j} $ contains multiple sound sources then $ c_{j} $ should have positive values for all occurred sound sources and zero for other sound classes that do not occur in $ s_{j} $. AudioSet does not provide labels that can be directly used as $ c_{j} $, for the reason that the label of audio clips are in 10-second level but not in anchor segment level. Instead, we use the audio tagging prediction of $ s_{j} $ from the trained SED system as the condition vector $ c_{j} $. This prediction can reflect the presence probability of sound events in $ s_{j} $. This proposed method does not require $ s_{j} $ to be clean. In training, we let the system to learn the following regressions:

\begin{equation} \label{eq1a}
f(s_{1} + s_{2}, c_{j}) \mapsto s_{j}
\end{equation}
\begin{equation} \label{eq1b}
f(s_{j}, c_{j}) \mapsto s_{j}
\end{equation}
\begin{equation} \label{eq1c}
f(s_{j}, c_{\neg j}) \mapsto \textbf{0}
\end{equation}

\noindent where $ j \in \{1, 2\} $. The symbol $ \neg j $ indicates any class index that is different from $ j $. Equation (\ref{eq1a}) represents learning from a mixture to a separated source $ s_{j} $ conditioned on $ c_{j} $. Equation (\ref{eq1b}) represents learning an identity mapping. That is, the system should learn to output the mixture itself if the condition describes the mixture. Equation (\ref{eq1c}) represents a zero mapping. That is, the system should output silence if it is conditioned on $ c_{\neg j} $. The vector $ \textbf{0} $ is an all zero vector. Fig. \ref{fig:framework} shows the framework of the proposed source separation with weakly labelled data method.

\subsection{Inference}
In inference, we first predict the presence or absence of sound classes in an audio clip using the SED system. Only the sound classes predicted as present in the audio clip will be separated. For the $ k $-th sound event, we use one-hot encoding as a conditional vector $ c_{k} = \{ 0, .., 0, 1, 0, ..., 0 \} $, where the $ k $-th index of $ c_{k} $ is 1 while others 0. The design of the condition vector is to control what sound class to separate. When only one value of $ c_{k} $ is non-zero indicates the system only output the separate source of one sound class. By choosing different condition vectors different separated sources can be obtained. Demos of the proposed system can are available\footnote{https://github.com/qiuqiangkong/audioset\textunderscore source\textunderscore separation}. 

\begin{figure*}[t]
  \centering
  \centerline{\includegraphics[width=\textwidth]{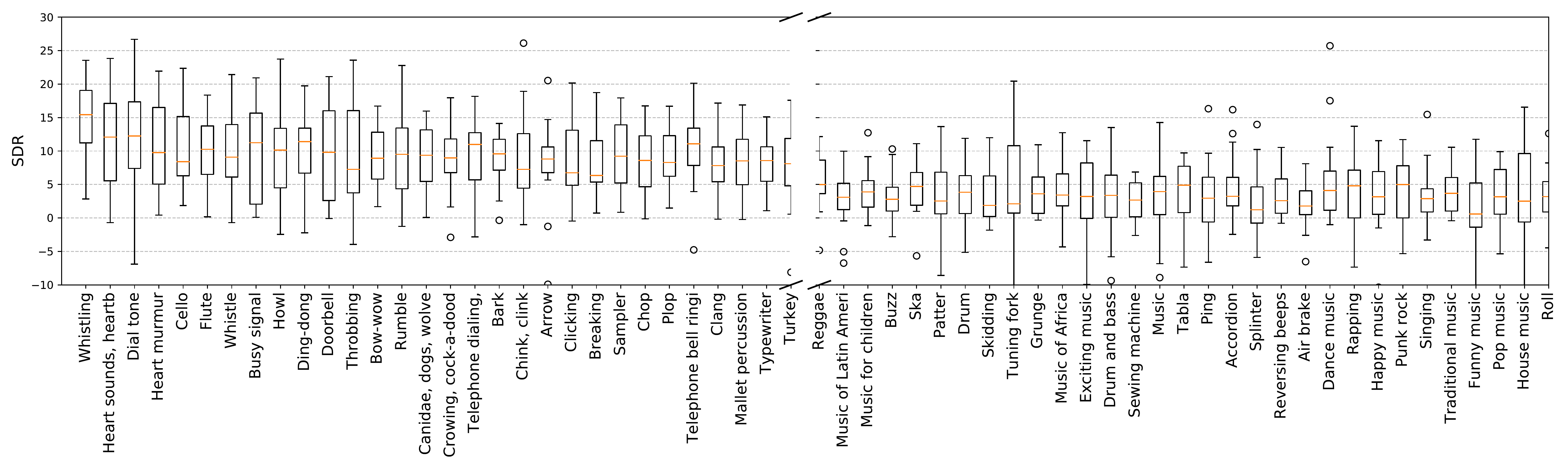}}
  \caption{SDRs of sound classes sorted by median value in a descending order.}
  \label{fig:sdrs}
\end{figure*}

\section{Experiments}\label{section:experiment}
We evaluate the the proposed source separation system on AudioSet. AudioSet is a large-scale audio dataset with an ontology of 527 sound classes \cite{gemmeke2017audio} in the released version. The balanced training subset contains 22,050 audio clips. The evaluation set consists of 20,371 audio clips. Most audio clips have a duration of 10 seconds. AudioSet is a weakly labelled dataset. All downloaded audio clips are converted to monophonic with a sampling rate of 32kHz. The SED system is trained on the full training set. Log mel spectrogram of audio clips are extracted with a STFT window of 1024 samples and a hop size of 320 frames which leads to 100 frames in a second. The number of mel frequency bins is set to 64 following \cite{kong2019cross}. The CNN for SED consists of 13 layers. A time distributed fully connected layer is applied to the final convolutional layer to obtain the presence probability of sound events. Adam optimizer \cite{kingma2014adam} with a learning rate of 0.001, $ \beta_{1} $ of 0.9 and $ \beta_{2} $ of 0.999 is used for training the SED system. In inference, the trained SED system is used to choose $ \tau $. The anchor segment has a duration of 1.6 seconds with $ \tau $ as the centre. 

For the source separation system, we apply a U-Net \cite{jansson2017singing}. The input to the U-Net is the spectrogram of the mixture of two anchor segments. The spectrogram is obtained by applying a STFT on the waveform with a window size of 1024 and a hop size of 256 in an anchor segment. The input has a shape of $ 160 \times 513 $ representing the number of frames and frequency bins. The U-Net consists of 4 encoder blocks and 4 decoder blocks. The number of feature maps in the encoder are 64, 128, 256 and 512 \cite{ronneberger2015u}. The number of feature maps in the decoder are 512, 256, 128 and 64. Each convolutional layer consists of a convolutional operation, a batch normalization and a ReLU non-linearity. The condition vectors is mapped to embedding vectors by a learnable matrix. The embedding vectors are added to after each ReLU operation in all layers as a bias. The condition vector controls what sources to separate. Adam optimizer with a learning rate of 0.001, $ \beta_{1} $ of 0.9 and $ \beta_{2} $ of 0.999 is used for training the source separation system. Mean absolute error between the estimated spectrogram $ \hat{S}_{j} $ and the target spectrogram $ S_{j} $ is used as the loss function in training the network.

In inference, we use the evaluation set in AudioSet for evaluation. we randomly select and mix two anchor segments of two sound classes under SDR of 0 dB. The task is to separate the anchor segments from the mixture. We evaluate the source separation system with source to distortion ratio (SDR), source to inferences ratio (SIR) and sources to artifact ratio (SAR) \cite{vincent2006performance}. The higher number indicates the better performance. Fig. \ref{fig:sdrs} shows the box plot of SDRs of sound classes sorted in a descending order. Most of sound classes achieve positive SDR indicating the effectiveness of the proposed source separation method. Some sound classes such as whistling achieves SDR of around 15 dB, dial tone of 12 dB and heart murmur of 10 dB. Fig. \ref{fig:sdrs} shows that the separation difficulty is different from class to class. The proposed system modeled with U-Net achieves a SDR, SIR and SAR of 5.67 dB, 9.38 dB and 11.15 dB, respectively. 

\section{Conclusion}\label{section:conclusion}
We have presented a source separation framework trained with weakly labelled data only. The proposed source separation system is a unified system trained on AudioSet which can separate a large amount of sound classes using a unified system. The proposed approach is a solution to the computational auditory scene analysis, and can detect and separate any sound sources in an audio recording. To begin with, a sound event detection system is trained to select anchor segments of sound classes in audio clips. Then, the mixture of two anchor segments and the condition vector of one anchor segment are used as input to the source separation system modeled by a U-Net. The proposed source separation system does not require clean sources for training. Overall, the system achieves an average SDR of 5.67 dB over 527 sound classes in AudioSet. In future, we will focus on improving the performance of the source separation system using weakly labelled AudioSet. 

\section{Acknowledgement}
This research was supported by EPSRC grant EP/N014111/1 ``Making Sense of Sounds'' and a Research Scholarship from the China Scholarship Council (CSC) No. 201406150082.

\ninept
\bibliographystyle{IEEEbib}
\bibliography{strings,refs}

\end{document}